\documentstyle[epsf]{article}
\newcounter{pepe}                         
{\end{eqnarray}%
\setcounter{equation}{\arabic{pepe}}%
} 

\newcommand{\Rb}{{\cal   R}}

\newcommand{\nne}{{\bf  e}}

\newcommand{\np}{{\bf   p}}       
\newcommand{\nq}{{\bf   q}}
\newcommand{\nr}{{\bf   r}}

\newcommand{\hr}{{\bf   \hat{r}}} 

\newcommand{\nOmega}{\mbox{\boldmath$\Omega$}} 
\newcommand{\hphi}{\hat{\mbox{\boldmath$\phi$}}}

\begin{document}
\mbox{} 
\vspace*{2.5\fill} 

{\Large\bf 
\begin{center}
%
Final-State Interaction in Electron Scattering  \\
by Polarized Nuclei\footnote{Talk presented by J.E. Amaro at the IVth
Workshop on Electromagnetically induced Two-Hadron Emission, Granada,
Spain, 1999}
\protect
%
\end{center}
} 

\vspace{1\fill} 

\begin{center}
{\large J.E. Amaro$^{a}$ and T.W. Donnelly$^b$} 
\end{center}

{\small 
\begin{center}
$^a$ {\em Departamento de F\'{\i}sica Moderna, 
          Universidad de Granada,
          Granada 18071, Spain}  \\   
$^b$ {\em Center for  Theoretical  Physics,
          Laboratory   for Nuclear Science  and  
          Dept. of Physics, }\\   
     {\em Massachusetts Institute  of  Technology, 
          Cambridge,  MA  02139, U.S.A.}\\[2mm] 
\end{center}
} 

\kern 1.5 cm 

\hrule \kern 3mm 

{\small 
\noindent
{\bf Abstract} 
\vspace{3mm} 

The cross section for the reaction 
$^{39}\vec{{\rm K}}(e,e'p)^{38}$Ar is computed, and
the dependence of the FSI on the initial
polarization angles of the nucleus $^{39}$K is analyzed.
The results are explained
in a  semi-classical picture of the reaction.
This procedure allows us to find
the best initial kinematical conditions for
minimizing the FSI.
 } 

\kern 3mm \hrule 


\section{Introduction}

 In this talk we explore the effects of the FSI in $(e,e'p)$ 
reactions using polarized nuclei. 
All of the measurements to date involving medium and heavy nuclei have 
been performed with unpolarized targets \cite{Bof96} 
and hence only the global
effects of FSI averaged over all polarization directions have been addressed
experimentally. Using instead polarized nuclei as targets, 
new possibilities to extract the full tri-dimensional momentum distribution 
of nuclei will become available \cite{Cab93,Cab94,Gar95,Cab95}.

The few theoretical studies 
of $(e,e'p)$ reactions involving polarized, medium and heavy nuclei in 
DWIA \cite{Bof88,Gar95,Ama98,Ama99} report 
 a dependence of FSI effects --- or nuclear transparency --- on the choice
of polarization angles. 
In  the present work  we show that these variations of the 
transparency can be understood in terms of the orientation of 
the initial-state nucleon's orbit
 and of the attenuation of the ejected  nucleon's flux.

We shall show  that one is able to predict 
the orientations of the target polarization that are optimal for 
minimizing the FSI effects,
providing the ideal situations for nuclear structure studies. 
This situation occurs when the nucleon is ejected directly away 
from the nuclear surface. On the other hand,  when the nucleon is ejected 
from the nuclear surface but in the opposite direction --- into the nucleus 
--- it has to cross the entire nucleus to exit on the opposite side, 
and the FSI effects are then found to be maximal.
This second situation is ideal for detailed studies of 
the  absorptive part of the FSI.
All of these situations 
can be selected simply by changing the direction of the nuclear polarization.


\section{Coincidence cross section 
         of polarized nuclei. Results}


Here we present the results of a DWIA calculation of 
the  $^{39}\vec{\rm K}(e,e'p)^{38}{\rm Ar}_{\rm g.s.}$
cross section in the extreme shell model. 
The present choice is  prototypical and
 the results can be generalized for any polarized
nucleus and can be addressed using more sophisticated nuclear models. 

We describe the ground state of $^{39}\vec{\rm K}$ as a hole in the 
$d_{3/2}$ shell of $^{40}$Ca. The initial nuclear state is 
100\% polarized in the direction $\Omega^*=(\theta^*,\phi^*)$.
\begin{equation}
|A(\Omega^*) \rangle = R(\Omega^*)
|d_{3/2}^{-1},m=\textstyle\frac32\rangle ,
\end{equation}
where $R(\Omega^*)$ is a rotation operator. In this simple model the
nuclear polarization is carried by the hole in the $d_{3/2}$ shell.
The polarization angles $(\theta^*,\phi^*)$ are the spherical
coordinates of the polarization vector $\nOmega^*$ with respect 
to the $\nq$-direction ($z$-axis) and with the $x$-axis in the 
scattering plane. 

The final hadronic state is given by a proton in the continuum with
energy $\epsilon'$ and momentum $\np'$, plus a daughter $A-1$ nucleus 
($^{38}$Ar) in the
ground state. This is described in the shell model as two holes in the
$d_{3/2}$ shell coupled to total spin $J=0$. 

The hole wave function is obtained by solving the Schr\"odinger
equation with a Woods-Saxon potential.
The wave function of the ejected proton is obtained by solving the
Schr\"odinger equation 
with an optical potential for positive energies. 

We compute the cross section as
\begin{equation}
\Sigma \equiv \frac{d\sigma}{dE'_ed\Omega'_ed\Omega'}
       = \sigma_M\left( v_L\Rb^L + v_T\Rb^T + 
                        v_{TL}\Rb^{TL} + v_{TT}\Rb^{TT} 
                 \right),
\end{equation} 
where $\sigma_M$ is the Mott cross section, $v_K$ are the electron 
kinematical factors given in \cite{Ras89}, and $\Rb^K$ are the
nuclear response functions.
See Refs. \cite{Ama98,Ama99} for more details of the model.

Next we show results of a calculation of the $(e,e'p)$ cross section
for different nuclear polarizations $(\theta^*,\phi^*)$. The
kinematics correspond to the quasi-elastic peak and in-plane emission
\[
q=500 \,{\rm Mev}/c,
\kern 5mm
\omega=133.5\, {\rm MeV},
\medskip
\kern 5mm
\phi=0,
\medskip
\kern 5mm
\theta_e = 30^{\rm o}
\]

\begin{figure}[hptb]
\begin{center}
\leavevmode
\mbox{\epsfxsize=8cm
\epsfbox[150 600 502 800]{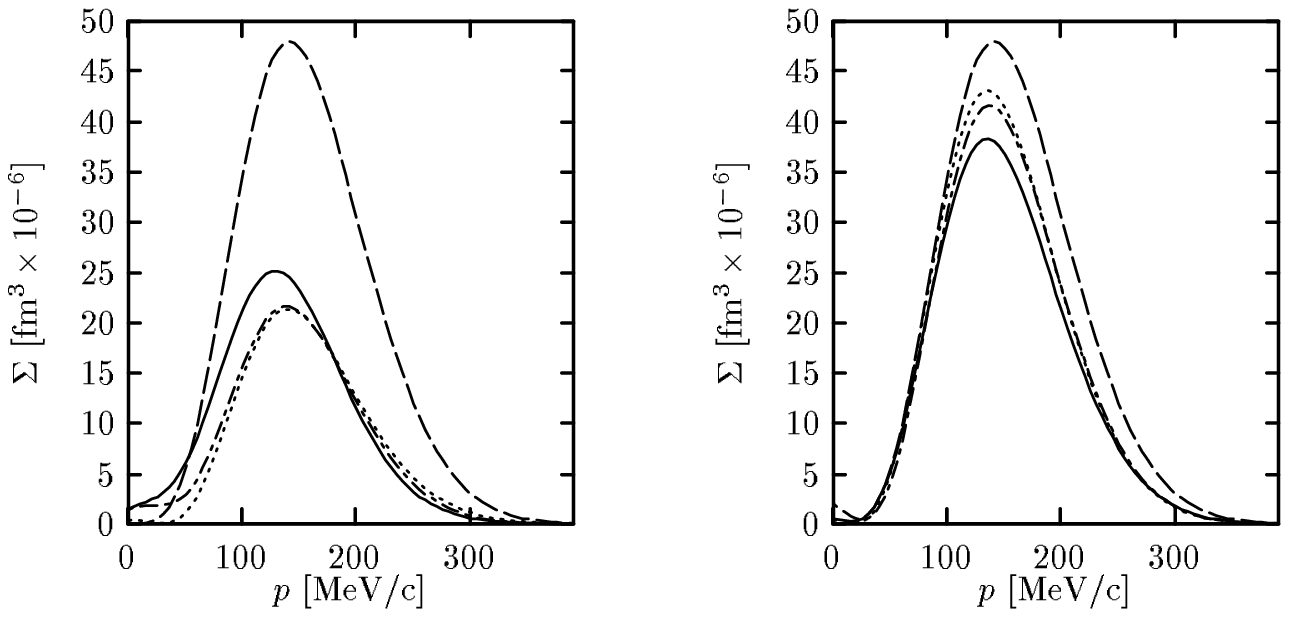}}
\end{center}
{\small \sf
{\bf Figure 1:} Cross section computed for two values of 
the nuclear polarization angles. Left: 
$\theta^*=90^{\rm o}$, $\phi^*=-90^{\rm o}$; 
right:
$\theta^*=90^{\rm o}$, $\phi^*=90^{\rm o}$.
 The meaning of the curves  
is the following: solid: DWIA; dashed: PWIA; dotted: DWIA 
but with just the imaginary part of the central optical potential;
dash-dotted: DWIA without spin-orbit contributions.
}
\end{figure}

In fig. 1 we show the cross section for 
$^{39}\vec{\rm K}$ polarized in the $-y$ direction,  (left)
and in the $y$ direction (right).
The solid lines are the full DWIA calculation,
using the optical potential of Schandt et al.
The dashed lines are the cross sections computed 
in PWIA, i.e., without FSI. The dotted lines correspond to the DWIA,
but including in the FSI just 
the central imaginary part of the optical potential,
while the dash-dotted lines include in addition the central real part
of the potential. 

Comparing the solid and dashed lines, we see that
 the effect of the FSI (solid lines relative to dashed lines) 
is quite dependent on the polarization of the nucleus. This fact suggest that
the ``transparency'' of the nucleus to proton propagation can be
maximized or minimized by selecting  a particular polarization of the
nucleus and that if one is able to understand physically the 
different behavior seen for the FSI effects in fig. 1, then it could be 
possible to make specific predictions about the reaction for
future experiments.


\section{A semi-classical picture of
         the reaction}


In order to understand physically the above results  
we will consider a 
semi-classical model of the reaction by assuming it to take 
place in two or more
steps as follows: first a proton with (missing) momentum $\np$ and energy
$\epsilon$ is knocked-out by the virtual photon and it acquires 
momentum $\np'$ and energy $\epsilon'=\epsilon+\omega$. Second, 
as this high-energy nucleon traverses the nucleus it undergoes 
elastic and inelastic scattering which, in our model, are
produced by the real and imaginary parts of the optical potential.

The important point here is that the nucleus is polarized in a 
specific direction. Accordingly, the initial-state nucleon 
can be localized in an oriented 
(quantum) orbit. From the knowledge of this orbit and of the missing momentum
one can predict the most probable location of 
the struck proton, and therefore one can specify the quantity
of nuclear matter that the proton must cross before
exiting from the nucleus with momentum $\np'$.

We illustrate the case of a particle in a $d_{3/2}$ wave.
which is the relevant state for our calculation.
First consider that the particle is
polarized in the $z$-direction ($\nOmega^* = \nne_3$).
The corresponding wave function can be written as
\begin{equation}
\textstyle
|\frac32\frac32\rangle 
= \psi_1 |\uparrow\rangle +\psi_2 |\downarrow\rangle .
\label{e11}
\end{equation}
where the up and down components are given by 
\begin{eqnarray}
\psi_1 &=& -\sqrt{\frac{3}{8\pi}}\sin\theta\cos\theta\;
           {\rm e}^{i\phi}R(r) 
           \label{e12}\\
\psi_2 &=& -\sqrt{\frac{3}{8\pi}}\sin^2\theta\;
           {\rm e}^{2i\phi}R(r).
            \label{e13}
\end{eqnarray}
Here the angles $(\theta,\phi)$ are the spherical coordinates of the 
particle's position $\nr$ and $R(r)$ is its radial wave function. 
The  spatial distribution is then given by the single-particle probability 
density
\begin{equation}
\rho(\nr) = |\psi_1|^2+|\psi_2|^2 =
    \frac{3}{8\pi}\sin^2\theta|R(r)|^2.
\label{e17}
\end{equation}

\begin{figure}
\begin{center}
\leavevmode
\epsfxsize=8cm
\epsfbox[50 100 400 550]{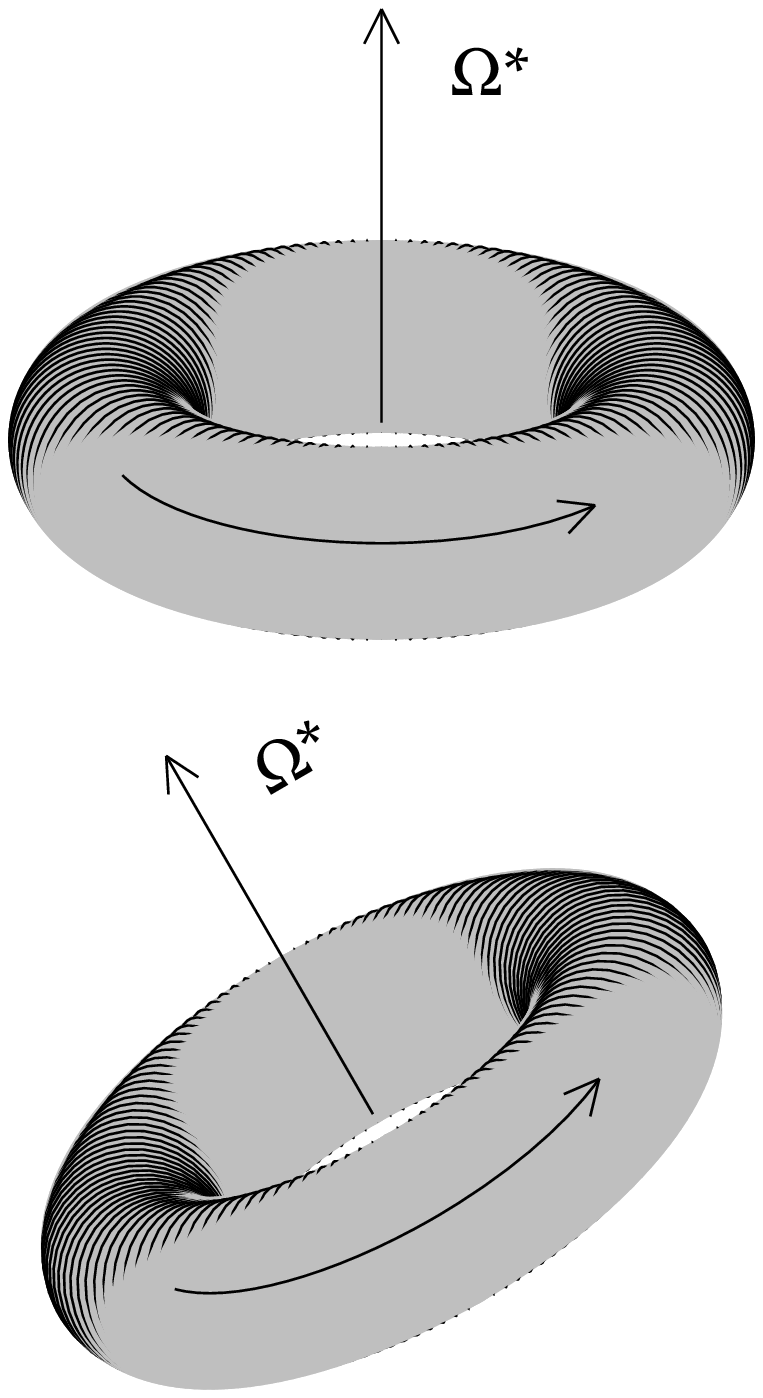}
\end{center}
{\small\sf 
{\bf Figure 2:} 
Pictorial representation of the spatial distribution of a proton in the
$d_{3/2}$  shell, shown as a torus-like distribution for 
two different  polarizations.
}
\end{figure}

Taking into account the form of the radial wave function for the $d_{3/2}$ 
wave, we can see that the particle is distributed around the center
of the nucleus in a toroidal-like (quantum) orbit as shown schematically in
fig.~2 (upper part).
In a semi-classical picture of the bound state, we can imagine the particle 
performing a rotatory orbit within the torus in a counter-clock
sense, as shown in the figure.
The shape of the distribution for arbitrary polarization $\nOmega^*$
is just a rotation of the above distribution, as also shown in
fig.~2 (bottom).

The next step is to localize the particle within  the orbit
for a  given value  of the missing momentum $\np$.
From elementary quantum mechanics 
we employ the
Fourier transform $\tilde{\psi}(\np)$ of the wave function 
and the position operator in momentum space $\hr=i\nabla_p$ to
define the local position of the nucleon in the orbit 
for momentum $\np$ in the following way:
\begin{equation}
\nr(\np) = \frac{{\rm Re}\; \tilde\psi^{\dagger}(\np)(i\nabla_p)
                            \tilde\psi(\np)}%
                {\tilde\psi^{\dagger}(\np)\tilde\psi(\np)} .
\end{equation}
This is a well-defined vector which represents the most probable 
location of a particle with momentum $\np$ when it is described
by a wave function $\psi$. Henceforth $\nr(\np)$  represents the position
of the particle in the orbit in the present semi-classical model.

For the case of interest here of the $d_{3/2}$ orbit polarized in the
 $z$-direction, we compute the position $\nr(\np)$ by using the wave
function given in eqs. (\ref{e11}--\ref{e13}) in momentum space:
\begin{equation}
{\rm Re}\, \tilde\psi^{\dagger}(\np)i\nabla_p\tilde\psi(\np)=  
-\frac{3}{8\pi}\frac{|\tilde{R}(p)|^2}{p}\sin\theta(1+\sin^2\theta)\hphi ,
\end{equation}
where now $(\theta,\phi)$ are the spherical coordinates of the missing
momentum $\np$, $\tilde{R}(p)$ is the radial wave function in momentum
space, and $\hphi$ is the unit vector in the azimuthal direction.
As we see, upon dividing by the momentum distribution 
(given by eq.  (\ref{e17}), but in momentum space) 
\begin{equation}
{\tilde\psi^{\dagger}(\np)\tilde\psi(\np)}= \frac{3}{8\pi}\sin^2\theta
|\tilde{R}(p)|^2 ,
\end{equation}
the radial dependence in the numerator and denominator goes away,
and we obtain an expectation value of position which is independent
of the radial wave function
\begin{equation}
\nr(\np) = -\frac{1+\sin^2\theta}{p\sin\theta}\hphi .
\end{equation}

This expression has been obtained for the polarization
in the $z$-direction. 
For a general polarization vector $\nOmega^*$ we just perform
a rotation of the vector $\nr(\np)$. Introducing the angle $\theta_p^*$
between $\np$ and $\nOmega^*$, 
we can write the nucleon position in a general way
\begin{equation}
\nr(\np)=-\frac{1+\sin^2\theta_p^*}{p^2\sin^2\theta_p^*}
         \,\nOmega^*\times\np.
         \label{e23}
\end{equation}

Using the above definitions we can give a 
physical interpretation of the results shown in fig. 1.
The kinematics 
for the case of the  $^{39}$K nucleus polarized
in the $-y$ direction 
are illustrated in fig.~3(a).
Therein, the momentum transfer points in the $z$-direction
and we show the missing-momentum vector $\np$ corresponding to the
maximum of the momentum distribution, $p\sim 140$ MeV/c. 
The momentum 
of the ejected proton $\np'$ is also shown in the picture.
For $\nOmega^*$ pointing in the $-y$ direction, the semi-classical orbit 
lies in the $xz$-plane and follows a counter-clockwise direction of rotation.
For these conditions, the most probable 
position of the proton before the interaction is indicated with a 
black dot near the bottom of the orbit. As the particle
is going up with momentum $\np'$ after the interaction with the
virtual photon, 
it has to cross all of the nucleus 
and exit it by the opposite side; thus
one expects that the FSI will be large in this situation, as shown
in the left  panel of fig.~1.
\begin{figure}[htb]
\begin{center}
\leavevmode
\epsfxsize=10cm \epsfbox{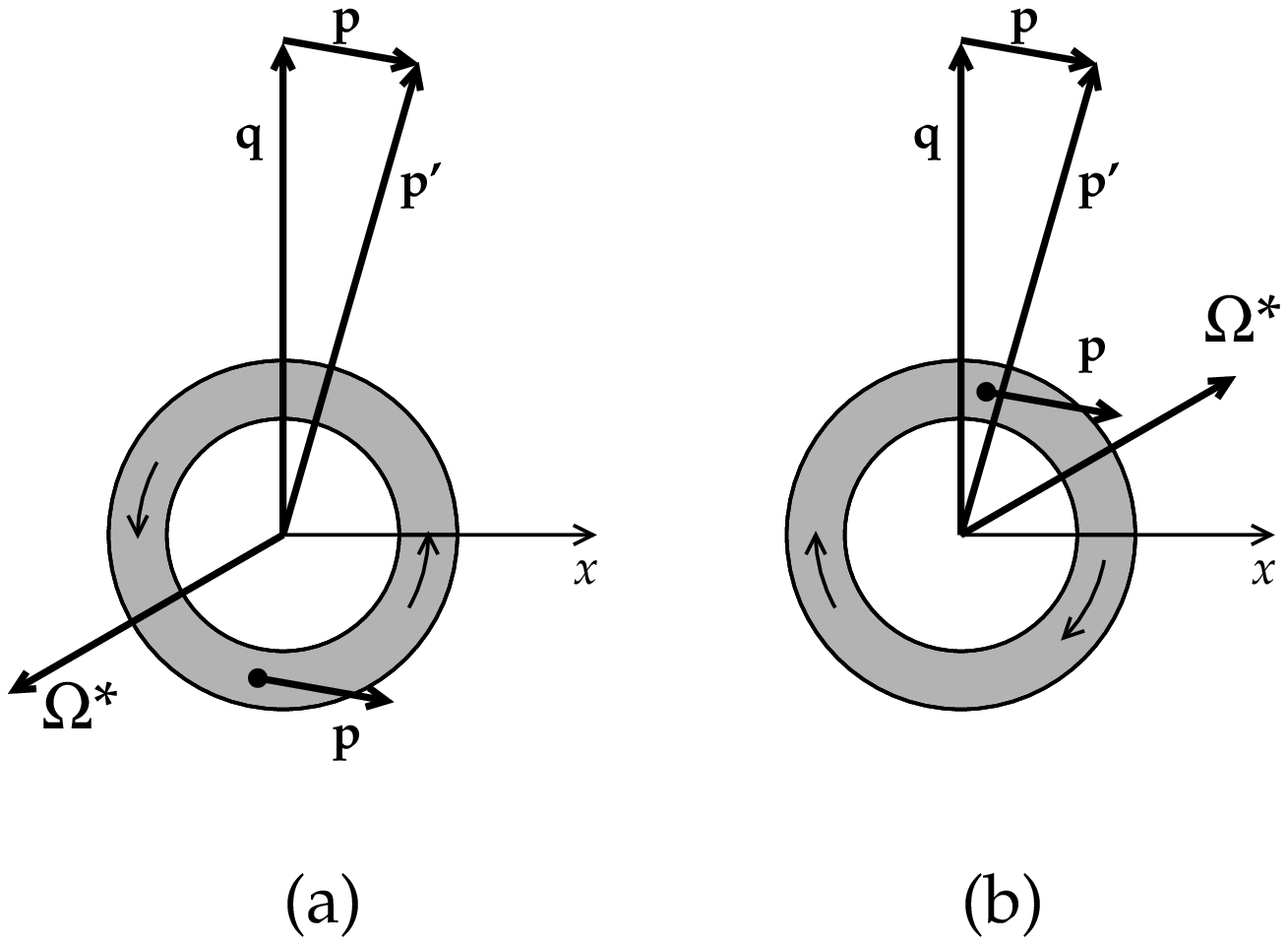}
\end{center}
{\small\sf  
{\bf Figure 3:} 
Semi-classical orbit and location of the proton for the
same  kinematics and  nuclear polarizations as in Fig. 1.
}
\end{figure}

In fig.~3(b) we show the picture for the opposite polarization
in the $y$-direction.
In this case the nucleon distribution in the orbit is the
same as in (a), but the rotation direction is the opposite,
Hence now it is more probable
for the nucleon to be located near the upper part of the orbit.
As the nucleon is still going up with the same momentum $\np'$,
the distance that it has to travel through the nucleus is much smaller
than in case (a), and hence one expects small FSI effects, namely, what is
seen in the right panel of fig.~1.

We have arrived at a very intuitive physical 
picture of why the FSI effects differ for different orientations of the 
nuclear spin: the polarization direction fixes the orientation of
the nucleon distribution (quantum orbit).
 For a given value of the missing momentum one can 
locate the particle in a definite position within the orbit, and therefore
within the nucleus. Assuming that the particle leaves the nucleus with
the known momentum $\np'$, one can  determine the quantity of
nuclear matter that it has to cross before exiting.

In order to check the above picture 
for any  nuclear polarization, we have 
computed the cross section for a set of 26 different nuclear polarization
angles expanding the $(\theta^*,\phi^*)$ plane. 
Using equation (\ref{e23}) we have computed 
 the distance $s$ of the 
nucleon  trajectory within the nucleus,
by choosing some appropriate
value for the nuclear radius $R$. A model of exponential
attenuation of the cross section due to nuclear absorption 
can be crafted in the following way:
\begin{equation}
\Sigma_{DWIA}\simeq \Sigma_{PWIA}\;{\rm e}^{-s/\lambda} ,
\end{equation}
where $\lambda$ is a free parameter to be interpreted as the 
mean free path (MFP). 
Within this  approximation, the nuclear transparency, defined as
the ratio between the DWIA and PWIA results, can be written as
\begin{equation}\label{e34}
T \equiv \frac{\Sigma_{DWIA}}{\Sigma_{PWIA}} \simeq {\rm e}^{-s/\lambda}.
\end{equation}

\begin{figure}[htbp]
\begin{center}
\leavevmode
\epsfxsize=10cm \epsfbox[80 550 600 800]{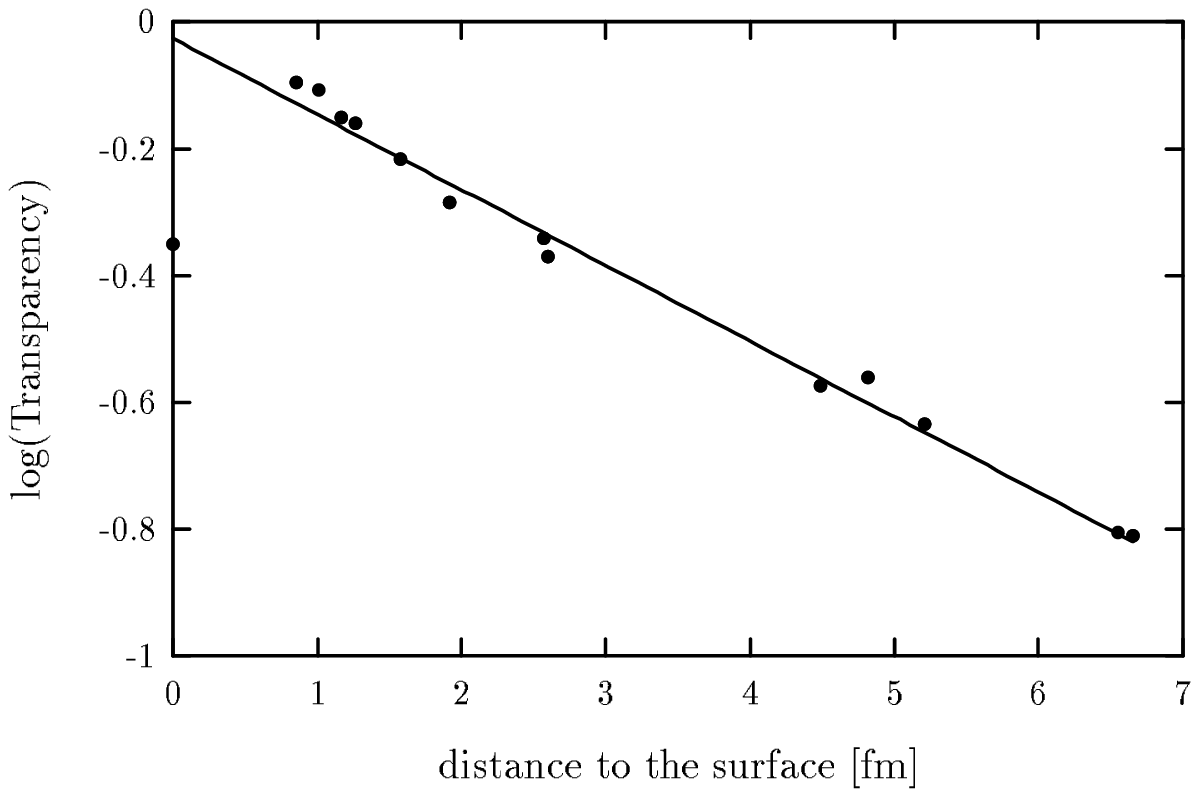}
\end{center}
{\small\sf  
{\bf Figure 4:} 
Nuclear transparency as a function of the nucleon path $s$ within the 
nucleus for different nuclear polarizations.
The FSI only include the imaginary part of the central optical 
potential. 
}
\end{figure}
In fig.~4 we show the nuclear transparency as a function of the
distance $s$ to the nuclear surface, computed  for different
polarizations, at the
maximum of the cross section. For the FSI we have used just the 
central, imaginary part of the optical potential.
In this figure we see that the dependence of $\log T$ can 
in fact be approximated by a straight line.
 By performing a linear regression we obtain a MFP of
$\lambda=8.4$ fm.  
This value 
is  quite independent of the radius $R$ in the
region between $r_{1/2}$ and $r_{1/10}$, where the nuclear density 
$\rho(r)$ takes the values $\rho(0)/2$ and $\rho(0)/10$, respectively.

\section{Applications to two-particle emission reactions}

Finally we give a possible application of the above model to 
two-hadron emission reactions. Consider $(e,e'N\pi)$ reactions 
from polarized nuclei in the $\Delta$-region.
By selecting the appropriate nuclear polarization, one could reduce or
enhance the FSI of the final $\Delta$ in the nuclear medium.
In fact, using the above model, one have control on the point of the
nucleus where the $\Delta$ is created. Making a crude estimation of
the length that the $\Delta$ travels before decaying
\begin{equation}
x \sim \frac{\hbar c}{\Gamma_{\Delta}} \sim \frac{200}{120}\,\rm fm
\sim 1.7\, fm
\end{equation}
we see that it could be possible to produce the $\Delta$ in the two
situations shown in figure 5.
\begin{figure}[htbp]
\begin{center}
\leavevmode
\epsfxsize=10cm 
\epsfbox[170 500 450 720]{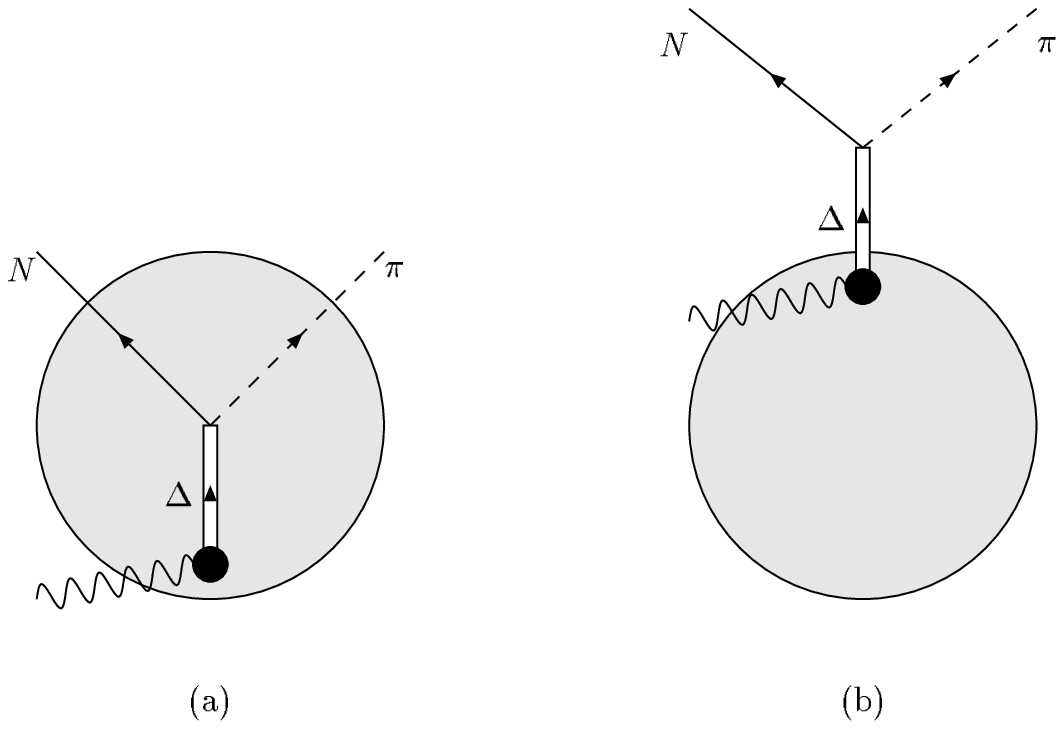}
\end{center}
{\small\sf  
{\bf Figure 5:} Electroproduction of a $\Delta$ from a polarized
nucleus for two different nuclear polarization directions. 
}
\end{figure}
In the case (a) the $\Delta$ is created near the nuclear surface 
and propagates into the nucleus. It has large FSI and decays inside
the nucleus into a pair $N+\pi$ which also interacts with the nucleus.
In the second case (b) the $\Delta$ propagates out of the nucleus. 
The FSI of the $\Delta$ is expected to be smaller. 
The $\Delta$ decays outside of the nucleus. This situation is cleaner
to study the $\Delta$ electroproduction amplitude in nuclei without
too much distortion by FSI. 
Case (a) is ideal to study the $\Delta$ properties in the nuclear
medium.

\section{Summary and conclusions}

We have studied  the reaction
$^{39}$K$(e,e'p)^{38}\rm Ar_{gs}$ for polarized $^{39}$K 
in DWIA.
We have studied the
dependence of the FSI as a function of the nuclear polarization
direction and introduced a physical  picture of the 
results in order to understand the different effects seen 
in the cross section. 

The argument to explain the FSI effects is based on the PWIA
and it has been illustrated by introducing the semi-classical  
concept of a nucleon orbit within the nucleus. 
For given kinematics
 we  can fix the 
 expectation value of the position of 
the nucleon within the nucleus before 
the interaction.  From this information we have computed the length
of the path that the nucleon travels across the nucleus
for each polarization.

Our results show that when the FSI effects are large the computed nucleon path 
through the nucleus is also large, whereas the opposite happens when the FSI 
effects are small. 
Thus, by selecting the appropriate nuclear polarization, one 
can  reduce or enhance  the FSI effects.
Such control should prove to be 
very useful in analyzing the results from future experiments with polarized 
nuclei. 

Finally, our model can also be applied to the $(e,e'N\pi)$ 
reaction in the $\Delta$ peak. 
Since by flipping the 
nuclear polarization one can go from big to small FSI effects of the 
$\Delta$ , this
opens the possibility of using this kind of reaction 
to distinguish the FSI effects   
from other issues of interest, such as the $\Delta$ electroproduction 
amplitudes in the medium.

\section*{Acknowledgments}

This work is supported  in part by funds provided
by the U.S.  Department of Energy (D.O.E.) under cooperative agreement
\#DE-FC01-94ER40818, in part  by DGICYT   (Spain) under Contract   No.
PB92-0927 and the Junta de Andaluc\'{\i}a  (Spain) and in part by NATO
Collaborative Research Grant \#940183.


\end{document}